\documentclass{aa}
\usepackage{graphicx}
\usepackage{epsfig}
\usepackage{natbib}
\bibpunct{(}{)}{;}{a}{}{,}
\begin{document}

\title{HD\,80606\,b, a planet on an extremely elongated orbit\thanks{
Based on observations made at the Observatoire de 
Haute-Provence (French CNRS) and at the W.M.~Keck Observatory, which 
is operated as a scientific partnership among the Californian 
Institute of Technology, the University of California and the 
National Aeronautics and Space Administration. The Observatory was 
made possible by the generous financial support from the W.M.~Keck 
Foundation.}}

\author{D.~Naef\inst{1} \and  D.W.~Latham\inst{2} \and 
M.~Mayor\inst{1} \and T.~Mazeh\inst{3} \and J.L.~Beuzit\inst{4} 
\and G.A.~Drukier\inst{3}\thanks{\emph{Present address}: Dept. of Astronomy, 
Yale University, P.O. Box 208101, New Haven, CT--06520-8101, USA}
 \and C.~Perrier-Bellet\inst{4} \and 
D.~Queloz\inst{1} \and J.P.~Sivan\inst{5} \and G.~Torres\inst{2} 
\and S.~Udry\inst{1} \and S.~Zucker\inst{3}}

\institute{Observatoire de Gen\`eve, 51 ch. des Maillettes, 
CH--1290 Sauverny, Switzerland \and 
Harvard-Smithsonian Center for Astrophysics, 60 Garden Street, 
Cambridge, MA--02138, USA \and 
School of Physics and Astronomy, Raymond and Beverly Sackler Faculty 
of Exact Sciences, Tel Aviv University, Tel Aviv 69978, Israel \and 
Laboratoire d'Astrophysique, Observatoire de Grenoble, 
Universit\'e J. Fourier, BP 53, F--38041 Grenoble, France \and 
Observatoire de Haute-Provence, F--04870 St-Michel L'Observatoire, 
France}

\offprints{Dominique Naef,
\email{dominique.naef@obs.unige.ch}}

\date{Received / Accepted}

\abstract{We report the detection of a planetary companion orbiting 
the solar-type star {\footnotesize HD\,80606}, 
the brighter component of a wide binary with a projected separation of
about 2000 AU. Using high-signal spectroscopic observations of the 
two components of the visual binary, we show that they are nearly identical. 
The planet has an orbital period of 111.8 days and a minimum mass of 
3.9\,${\mathrm M_{\rm Jup}}$. With $e$\,=\,0.927, this planet has 
the highest orbital eccentricity among the extrasolar planets detected so far. 
We finally list several processes this extreme eccentricity could result from. 
}

\titlerunning{An eccentric extrasolar planet orbiting HD 80606}
\authorrunning{D.~Naef et al.}
\maketitle

\section{Introduction}\label{intro}

We report in this paper on our radial-velocity measurements of 
{\footnotesize HD}\,80606, the primary star of the visual binary 
system {\footnotesize HD}\,80606--{\footnotesize HD}\,80607. These 
observations reveal the presence of a 3.9\,Jovian-mass planet 
(minimum mass) in a very eccentric orbit around this solar-type star.

The variable velocity of {\footnotesize HD}\,80606 was first noticed 
by the {\sl G--Dwarf Planet Search} \citep{Latham00}, a 
reconnaissance of nearly 1000 nearby G dwarfs that uses the 
{\footnotesize HIRES} high-resolution spectrograph \citep{Vogthires} 
mounted on the 10-m Keck~1 telescope at the W.M.~Keck Observatory 
(Hawaii, USA) to identify extrasolar planet candidates. 
The star was then followed up by the 
{\sl {\footnotesize ELODIE} Planet Search Survey} team 
\citep{mayorpp400,Udryman} using the {\footnotesize ELODIE} 
fiber-fed echelle spectrograph \citep{Baranne96} mounted on the 
Cassegrain focus of the 1.93--m telescope at the Observatoire de 
Haute-Provence ({\footnotesize CNRS}, France).

The {\footnotesize ELODIE} velocities are obtained by 
cross-correlating the observed spectra with a numerical template. 
The instrumental drifts are monitored and corrected using the 
"simultaneous Thorium-Argon technique" with dual fibers 
\citep{Baranne96}. The achieved precision with this instrument is 
of the order of 10\,m\,s$^{\rm -1}$. The {\footnotesize HIRES} 
instrumental profile and drifts are monitored using an Iodine gas 
absorption cell \citep{marcy92}. The radial velocities are derived 
from the spectra using the {\footnotesize TODCOR} code 
\citep{zucker94}, a two-dimensional correlation algorithm.

The observations of {\footnotesize HD}\,80606 started in April 1999 
with {\footnotesize HIRES}. With a velocity difference of 
267\,m\,s$^{\rm -1}$ in less than one month between the first two 
measurements, the variability of this source was quickly detected. 
In July 1999, we started an {\footnotesize ELODIE} 
radial-velocity follow up of 6 non-active slow-rotating radial-velocity variable 
stars detected with {\footnotesize HIRES}, including {\footnotesize HD}\,80606. 
The first {\footnotesize ELODIE} measurement for this star was obtained 
during our November 1999 run.
The discovery of the planetary companion orbiting 
{\footnotesize HD}\,80606 has been recently announced together with 
10 other new extrasolar planet candidates (April 4th 2001 {\footnotesize ESO PR}\footnote{www.eso.org/outreach/press-rel/pr-2001/pr-07-01.html}).
Among these is the planetary companion to {\object{HD\,178911\,B} 
(Zucker et al. in prep.), another one of the candidates identified 
with {\footnotesize HIRES}. 

The stellar characteristics of the two components of the 
{\footnotesize HD}\,80606--{\footnotesize HD}\,80607 visual binary 
are presented in Sect.\,\ref{stars}. The radial-velocity data and 
the orbital solution are presented in Sect.\,\ref{RVs}. The very 
high orbital eccentricity is discussed in Sect.\,\ref{disc}.
The 61 radial-velocity measurements presented in Sect.\,\ref{RVs} as well as the 
Iron line list we used in Sect.\,\ref{stars} will be made available in electronic 
form at the {\footnotesize CDS} in Strasbourg.

\section{Stellar properties} \label{stars}

{\footnotesize HD}\,80606 (\object{HIP\,45982}) and 
{\footnotesize HD}\,80607 (\object{HIP\,45983}) are the two 
components of a visual binary system. They have common proper 
motions and the fitted systemic velocity for {\footnotesize HD}\,80606 
($\gamma$\,=\,3.767\,$\pm$\,0.010\,km\,s$^{\rm -1}$) is almost equal to the
mean radial velocity measured for {\footnotesize HD}\,80607
($\langle RV\rangle$\,=\,3.438\,$\pm$\,0.025\,km\,s$^{\rm -1}$). The difference between the 
two values can be explained by the binary orbital motion.
The main stellar characteristics of {\footnotesize HD}\,80606 and 
{\footnotesize HD}\,80607 are listed in Table\,\ref{starparam}. The 
spectral types, apparent magnitudes, colour indexes, parallaxes and 
proper motions are from the HIPPARCOS Catalogue \citep{ESA97}.
The projected stellar rotational velocity, $v\sin i$, was measured 
using the mean {\footnotesize ELODIE} cross-correlation dip width 
and the calibration by \citet{Queloz98}. The rms of the 
{\footnotesize HIPPARCOS} photometric data is large for both stars 
($\sigma_{H_{p}}$\,$\simeq$\,40\,mmag) but this measured
scatter is classified as {\sl `duplicity--induced--variability'} in this 
catalogue. The angular separation between the two visual components is about 
30\,\arcsec. This value is not much larger than the satellite detector size so 
contamination from one component onto the other is probably responsible for the 
observed scatter. The contamination is also probably responsible for 
the difference in parallaxes (a factor of two) and for the 
abnormally large uncertainties on this parameter 
($\sigma_{\pi}$\,$\simeq$\,1\,mas is expected 
with {\footnotesize HIPPARCOS} for a 9th magnitude star).

We derived the atmospheric parameters ({\footnotesize LTE} analysis) 
using {\footnotesize HIRES} high signal--to--noise spectra with the 
same method as in \citet{Santosmet}. We used the same line list and 
oscillator strengths as these authors, except for some lines that 
could not be used because they were out of the {\footnotesize HIRES} 
spectral coverage or fell just between two non-overlapping orders of 
the echelle spectra. Our line list finally consisted of 
18 \ion{Fe}{i} lines and only 3 \ion{Fe}{ii} lines. 
We estimated the uncertainties on the derived atmospheric parameters 
in the same way as in \citet{GV}. The two stars have almost the 
same Iron abundance and are very metal-rich dwarfs (respectively 2.7 and 2.4 
times the solar Iron abundance). An independent study 
(Buchhave et al. in prep.) using the same {\footnotesize HIRES} spectra but a 
different line list gives consistent results.
For the Lithium abundance measurement, we summed all our 
{\footnotesize ELODIE} spectra in the 
\mbox{$\lambda$ 6707.8 \AA\ \ion{Li}{i}} line region. No trace of 
Lithium was detected giving upper limits (3--$\sigma$ confidence 
level) on the corresponding equivalent widths for both stars. The 
abundance upper limits were then derived using the curves of growth 
by \citet{Soder93}. The Lithium abundances are scaled with 
$\log$\,$n({\rm H})$\,=\,12.
 
\begin{table}[t!]
\caption{
\label{starparam}
Observed and inferred stellar parameters for HD\,80606 and HD\,80607}
\begin{tabular}{llr@{  $\,\pm\,$  }lr@{  $\,\pm\,$  }l}
\hline
\multicolumn{2}{c}{}        & \multicolumn{2}{c}{{\footnotesize HD}\,80606}  & \multicolumn{2}{c}{{\footnotesize HD}\,80607}\\
\hline
Sp.\,Type                   &                     & \multicolumn{2}{c}{G5}       & \multicolumn{2}{c}{G5}\\
$m_{\rm V}$                 &                     & 9.06  & 0.04                 & 9.17  & 0.04\\
$B-V$                       &                     & 0.765 & 0.025                & 0.828 & 0.029\\
$\pi$                       & (mas)               & 17.13 & 5.77                 & 9.51  & 8.76\\
Distance                    & (pc)                & 58.4  & $^{29.6}_{14.7}$     & 105   & $^{1228}_{51}$\\
$\mu _{\alpha}\cos(\delta)$ & (mas yr$^{\rm -1}$) & 46.98 & 6.32                 & 42.90 & 9.23\\
$\mu _{\delta}$             & (mas yr$^{\rm -1}$) & 6.92  & 3.99                 & 8.26  & 5.88\\
$T_{\rm eff}$               & ($\degr$K)          & 5645  & 45                   & 5555  & 45\\
$\log g$                    & (cgs)               & 4.50  & 0.20                 & 4.52  & 0.15\\
$\xi _{\rm t}$              & (km\,s$^{-1}$)      & 0.81  & 0.12                 & 0.91  & 0.12\\
$[$Fe/H$]$                  &                     & 0.43  & 0.06                 & 0.38  & 0.06 \\
$v\sin i$                   & (km\,s$^{-1}$)      & 0.9   & 0.6                  & 1.4   & 0.4\\
$W_{\lambda,{\rm Li}}$      & (m\AA)              &\multicolumn{2}{c}{$<$\,2.5}  & \multicolumn{2}{c}{$<$\,3.0}\\
$\log$\,$n({\rm Li})$       &                     &\multicolumn{2}{c}{$<$\,0.78} & \multicolumn{2}{c}{$<$\,0.77}\\
\hline
\end{tabular}
\end{table}

\section{Radial-velocity  analysis and orbital solution}\label{RVs}

On the 24th of April 2001, we had in hand a total of 61 
radial-velocity measurements for analysis: 6 from 
{\footnotesize HIRES} and 55 from {\footnotesize ELODIE}. 
The mean uncertainty on the velocities are of the order of 
14\,m\,s$^{\rm -1}$ (systematic error + photon noise) for both 
instruments. The {\footnotesize HIRES} velocities have an arbitrary 
zero point. From contemporaneous observations, we applied a preliminary 
shift to these velocitites to bring them to the 
{\footnotesize ELODIE} system: 
$\Delta$RV\,=\,$+$3.807\,km\,s$^{\rm -1}$. 
To account for possible errors in this zero-order shift, the orbital 
solution presented in Table\,\ref{orbelem} includes the residual 
velocity offset $\Delta {\rm RV}_{\rm H-E}$ between 
{\footnotesize HIRES} and {\footnotesize ELODIE} as an additional 
free parameter. The obtained $\Delta {\rm RV}_{\rm H-E}$ is 
consistent with zero. 
Figure\,\ref{orbs}a shows the temporal 
velocities for {\footnotesize HD}\,80606. The phase-folded 
velocities are displayed in Fig\,\ref{orbs}c. 
The fitted orbital eccentricity is extremely high --- 
$e$\,=\,0.927\,$\pm$\,0.012.
Assuming a mass of 1.1\,${\mathrm M_{\odot}}$ for 
{\footnotesize HD}\,80606, a typical value for a very 
metal-rich star with a solar effective temperature, the planetary 
companion minimum mass is 
$m_{\rm 2}$\,=\,3.90\,$\pm$\,0.09\,${\mathrm M_{\rm Jup}}$. 
The semimajor axis is 0.469\,AU and the orbital separation ranges 
from 0.034\,AU (periastron) to 0.905\,AU (apastron). 

\begin{table}[t!]
\caption{\label{orbelem} Fitted orbital elements to the 
radial-velocity measurements for HD\,80606. The velocities obtained 
with the HIRES spectrograph (H) have been set into the ELODIE (E) 
system}
\begin{tabular}{llr@{  \,$\pm$\,  }l}
\hline
$P$                         & days                           & 111.81         & 0.23\\
$T$                         & HJD                            & 2\,451\,973.72 & 0.29\\
$e$                         &                                & 0.927          & 0.012\\
$\gamma$                    & km\,s$^{\rm -1}$               & 3.767          & 0.010\\
$w$                         & $\degr$                        & 291.0          & 6.7\\ 
$K_{\rm 1}$                 & m\,s$^{\rm -1}$                & 411            & 31\\
$\Delta {\rm RV}_{\rm H-E}$ & m\,s$^{\rm -1}$		     & 1.5            & 8.5\\
$a_{\rm 1} \sin i$          & $10^{\rm -3}$AU                & 1.581          & 0.037\\
$f_{\rm 1}(m)$              & $10^{-8}$${\mathrm M_{\odot}}$ & 4.26           & 0.29\\
$m_{\rm 2} \sin i$          & ${\mathrm M_{\rm Jup}}$        & 3.90           & 0.09\\
$N$                         &                                & \multicolumn{2}{c}{55(E) + 6(H)}\\ 
$\sigma_{\rm O-C}$          & m\,s$^{\rm -1}$                & \multicolumn{2}{c}{17.7 (E:16.3, H:29.9)}\\
\hline
\end{tabular}
\end{table}

The residuals to the fitted orbit cannot be explained by our measurement errors. 
The computed $\chi^2$ probability for the full set of data is lower than 10$^{-3}$ 
($\chi^2$\,=\,102.45, $\nu$\,= number of degrees of freedom 
=\,$N-$7 free parameters\,=\,54).
Our measurement errors are correctly estimated for both instruments 
(see e.g. the low residuals value obtained for {\footnotesize HD}\,178911\,B, 
Zucker et al. in prep.). The very low $P(\chi^{\rm 2})$ 
value found for {\footnotesize HD}\,80606 can therefore not result 
from an underestimation of our measurement errors. 
Using our {\footnotesize HIRES} high-signal spectrum, no chromospheric emission 
is detected for {\footnotesize HD}\,80606 so the expected stellar jitter is low 
\citep[a few m\,s$^{-1}$, see e.g.][]{Santosact,Saar}.
Activity related processes are therefore probably not responsible for the 
observed residuals. The later could be explained by the presence of another planet 
around {\footnotesize HD}\,80606 on a longer period orbit perturbating the stellar
radial-velocity signal induced by the inner companion. No clear velocity 
trend was detected from the residuals curve 
(see Fig.\,\ref{orbs}b). Future measurements should help 
to solve the question.

\section{Discussion}\label{disc}

\begin{figure}[t!]
    \epsfig{figure=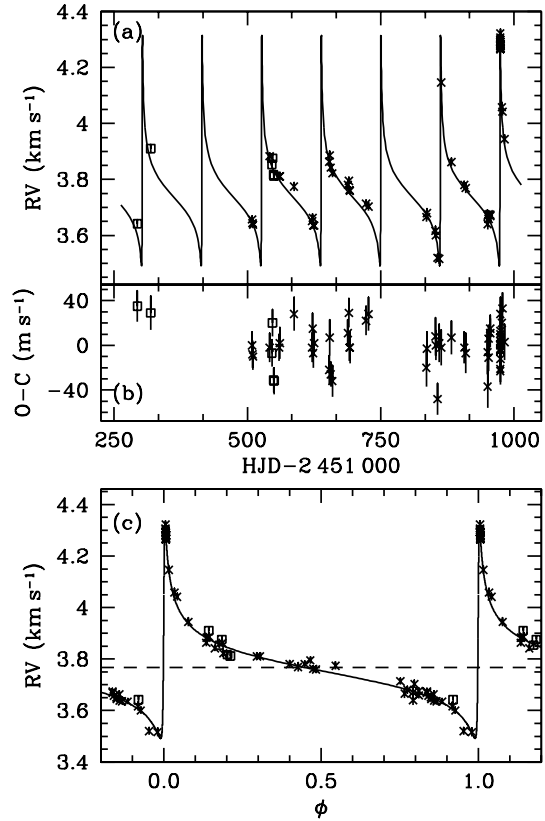,width=0.8\hsize}
    \caption{\label{orbs} HD 80606 radial-velocity data. 
    Crosses: Elodie-OHP measurements. 
    Open squares: Hires-KECK measurements. 
    {\bf a)}\,Temporal velocities. 
    {\bf b)}\,Residuals around the solution.
    {\bf c)}\,Phase-folded velocities}
\end{figure}

The fitted orbital eccentricity is the
highest found so far for an extrasolar planet orbiting a solar-type
star. The orbital eccentricities for extrasolar planets with period
longer than 100\,days almost cover the full possible range
\citep{Mayorten, Udryastrbio}: from nearly circular (see e.g. the
recently announced planet around {\footnotesize HD}28185,
$P$\,=\,385\,days, $e$\,=\,0.06, {\footnotesize ESO PR}\footnote{www.eso.org/outreach/press-rel/pr-2001/pr-07-01.html}) 
to nearly unity, as in the case of {\footnotesize HD}\,80606. The
distribution of the eccentricities of the planetary orbits might be a
keystone in understanding the formation processes of planets, as
was pointed out early in the study of extrasolar planets by
\citet{Mazeh97}.

Before discussing any mechanism that could have generated the 
eccentricity of {\footnotesize HD}\,80606, it is interesting to note
that the eccentricity distribution of the planets with long orbital 
periods found so far is strikingly similar to that of the binary 
orbits \citep{Heacox,Step2000,Step2001,Mayorten,Mazeh2000}. In particular, the high 
eccentricity of {\footnotesize HD}\,80606 is very similar to one of the highest 
eccentricity found so far for a spectroscopic binary --- $0.975$
\citep{Duquennoy}. The similarity of the two eccentricity distributions
does {\it not} prove that the planets and the low-mass stellar
companions come from the same population. The large gap between
the mass distribution of the planets and that of the stellar companions 
\citep{Jorissen,zucker2001} and the differences in the metallicity distributions for stars 
with and without planets \citep{Santosmet2} strongly suggests that we are dealing 
with two distinct populations. Nevertheless, we might need to 
look for mechanism(s) that can produce a range of eccentricities from zero up to 
unity for the two populations.

A mechanism to generate eccentric orbits could be the gravitational
interaction of a planet (and a binary) with a disk 
\citep{Arty91,Arty92}. However, a recent study \citep{Papaloizou} 
suggests that for a standard disk model this can happen
only for massive companions, at least in the range of brown
dwarf masses. For companions with planetary masses the disk probably acts
to damp the eccentricity growth, and therefore can not explain the
observed high eccentricities.

Another possible mechanism is the gravitational interaction with
another planet(s). This could be via dynamical instability 
\citep[e.g. ][]{Weiden, Rasio, Lin, Ford} or through some resonant 
interaction with a disk and another planet 
\citep{Murray}. The instabilities naturally lead to high
eccentricities, specially if they involve ejection of another planet
out of the system. The resonant interaction, on the other hand, seems
to need some fine tuning for generating eccentricities as high as the
one found here.

A possible clue to the origin of the particularly high eccentricity
found here could have been found in the fact that {\footnotesize
HD}\,80606 resides in a stellar wide binary. At least one other
planet, the one orbiting \object{16\,Cyg\,B}, was found with a high eccentricity
in a wide binary. A few studies 
\citep{Mazeh16cyg, Holman} have suggested that the
high eccentricity of 16\,Cyg\,B is because of the gravitational
interaction with the distant stellar companion. In this model, the
{\sl 'tidal`} interaction of the distant companion induces an eccentricity
modulation into the planetary orbit on a long timescale. The present
phase of the cycle is close to the highest point of the eccentricity
modulation.

However, this interpretation does not apply here. This is so because
the modulation timescale induced by {\footnotesize HD}\,80607 is of 
the order of 1\,Gyr \citep[e.g. ][]{Mazeh79}. This is very long 
relative to the relativistic periastron passage modulation, which 
is of the order of 1\,Myr. The precession of the longitude of the 
periastron induced by the relativistic effect completely suppresses 
the third-body modulation.
To hold on to the third-body interpretation we have to {\sl assume} 
an additional body in the system, in an orbit around 
{\footnotesize HD}\,80606 with a period of the order of 100 yrs. The 
present radial-velocity measurements can not rule out such a 
companion. To be consistent, this model has to apply
for all the planets with high eccentricity, above, say, 0.6 --- an
eccentricity that does not seem to be that rare anymore. Note also 
that this model requires a large angle between the plane of motion 
of the planet and that of the perturbating body.

In short, as stressed by \citet{Mayormicro} and \citet{Step2001}, we need a 
consistent model that will account for the distribution of eccentricities of the 
planetary orbits and for its similarity to the distribution for stellar companions. 
It seems that further observations and theoretical work are needed to reach a 
consensus about such a model.

\begin{acknowledgements}
We acknowledge support from the Swiss National Research Found 
({\footnotesize FNRS}), the Geneva University and the French 
{\footnotesize CNRS}. We are grateful to the Observatoire de 
Haute-Provence for the generous time allocation. This work was supported 
by the US-Israel Binational Science Foundation through grant 97-00460 and 
the Israeli Science Foundation (grant no. 40/00). We give special 
thanks to Yves Debernardi from Institut d'Astronomie de Lausanne 
for additional {\footnotesize ELODIE} radial-velocity measurements 
obtained during his own observing run. This research has made use 
of the {\footnotesize SIMBAD} database, operated at 
{\footnotesize CDS}, Strasbourg, France.
\end{acknowledgements}

\bibliographystyle{apj}
\bibliography{De293}
\end{document}